\documentclass{kluwer}    % Specifies the document style.
\usepackage{graphics,lscape}

\newdisplay{guess}{Conjecture}

\begin{document}                                                                                   
\begin{article}
\begin{opening}         
\title{A 2.09-h Photometric Periodicity in GW Librae\thanks{This paper uses observations
made from the South African Astronomical Observatory (SAAO).}} 
\author{Patrick A. \surname{Woudt} and Brian \surname{Warner}}  
\runningauthor{Patrick A. Woudt and Brian Warner}
\runningtitle{A 2.09-h Photometric Periodicity in GW Librae}
\institute{Dept. of Astronomy, University of Cape Town, Rondebosch 7700, South Africa\\
E-mail: pwoudt@artemisia.ast.uct.ac.za, warner@physci.uct.ac.za}
\date{15 October 2001}

\begin{abstract}
We have found a 2.09 h modulation in brightness of the dwarf nova GW Lib. This bears
no special relationship to the 1.28 h spectroscopic period that is believed to be the
orbital period. It was present in May 2001, but not in observations made in 1997 and 1998
and is of unknown origin. Similar unexplained periodicities are present in FS Aur and V2051 Oph.
\end{abstract}

\end{opening}           

\section{Introduction}  

 GW Lib was discovered in 1983 as a ninth magnitude star (Gonz\'alez 1983). It later 
faded to $m_v \sim 18.5$, on the basis of which (and in the absence of any spectroscopic 
observations) it was classified as a nova. Spectra obtained at minimum, however, show 
the emission lines and broad hydrogen absorption lines characteristic of a dwarf nova 
with very low rate of mass transfer (Duerbeck \& Seitter 1987; Ringwald, Naylor \& Mukai 1996; 
Szkody, Desai \& Hoard 2000). As a result, it is clear that GW Lib is a member of the SU 
UMa class of dwarf novae, with very long interval between outbursts (see Warner 
1995a,b for reviews of this class of cataclysmic variable star). No outburst other than that 
of 1983 has been observed; the amplitude of that suggests a superoutburst. GW Lib is 
therefore probably a system like WZ Sge.
    
 Interest in GW Lib has been intensified by the discovery that its white dwarf primary 
has non-radial oscillations in the manner of a ZZ Cet variable (Warner \& van Zyl 1998; 
van Zyl et al. 2000). The oscillations appear as brightness variations with maximum range  
$\sim$ 0.03 mag and power in the regions of 236, 376 and 650 s. The spectral continuum and 
absorption line profiles give an effective temperature of $\sim$ 11\,000 K (Szkody, Desai \& 
Hoard 2000), which is consistent with the white dwarf component being in the ZZ Cet 
instability strip.
    
 Extensive high speed photometric observations were obtained in March, April and 
September 1997 and in May 1998 (van Zyl et al. 2000). In none of these was any 
persistent brightness modulation detected that could be ascribed to an orbital period. In 
particular, neither the spectroscopic period of 79.4 min obtained by Szkody, Desai \& Hoard,
nor the improved spectroscopic period of 76.78 min found by Thorstensen et al (2001), was present
in the photometry.
     
 In order to check that GW Lib is still behaving as in former years, we sampled the 
light curve in May 2001. The ZZ Cet oscillations were still prominently present, but in 
addition there was a clear periodic modulation with period near 2 h which encouraged us 
to investigate further.

\section{Observations and Analysis}

 Our observing runs are listed in Table 1. Observations taken during the first week
of observing were made with the 40-inch reflector at the Sutherland site of the South African Astronomical 
Observatory; the following week we moved to the 74-inch reflector. In both cases the 
University of Cape Town CCD photometer (O'Donoghue 1995) was attached to the 
telescope. In order to maximize the signal, observations were made in `white light'.

\begin{table}
 \centering
  \caption{Observing log.}
  \begin{tabular}{@{}ccccccc@{}}
\hline
 Run No.  & Date of obs.          & HJD of first obs. & Length    & $t_{in}$ & Tel.    & $<$V$>$ \\ 
          & (start of night)      &  (+2452000.0)     & (h)       &     (s)   &        & (mag)   \\[10pt]
\hline
 S6217    & 18 May 2001           &         48.40493  &   4.88    &      20   &  40-in &  16.9    \\
 S6219    & 20 May 2001           &         50.54336  &   2.49    &      20   &  40-in &  16.8    \\
 S6221    & 21 May 2001           &         51.27101  &   4.98    &      20   &  40-in &  17.0    \\
 S6227    & 23 May 2001           &         53.40757  &   3.51    &      15   &  74-in &  16.9    \\
 S6230    & 26 May 2001           &         56.41459  &   5.29    &      20   &  74-in &  17.0    \\
 \hline
 \end{tabular}
\label{tab1}
\end{table}

 The individual light curves are displayed in Figure 1 and an average light curve, with 
the mean brightness subtracted, is shown in Figure 2. In both Figures the abscissa is the 
phase in the 2.091 h period deduced below. Most of the rapid variation is due to the ZZ 
Cet pulsations. The continuous presence of a double peaked modulation with a period 
near 2 h and a peak-to-peak range of 0.10 mag is quite evident.
    
\begin{figure}
\centering
 \resizebox{\hsize}{!}{\includegraphics{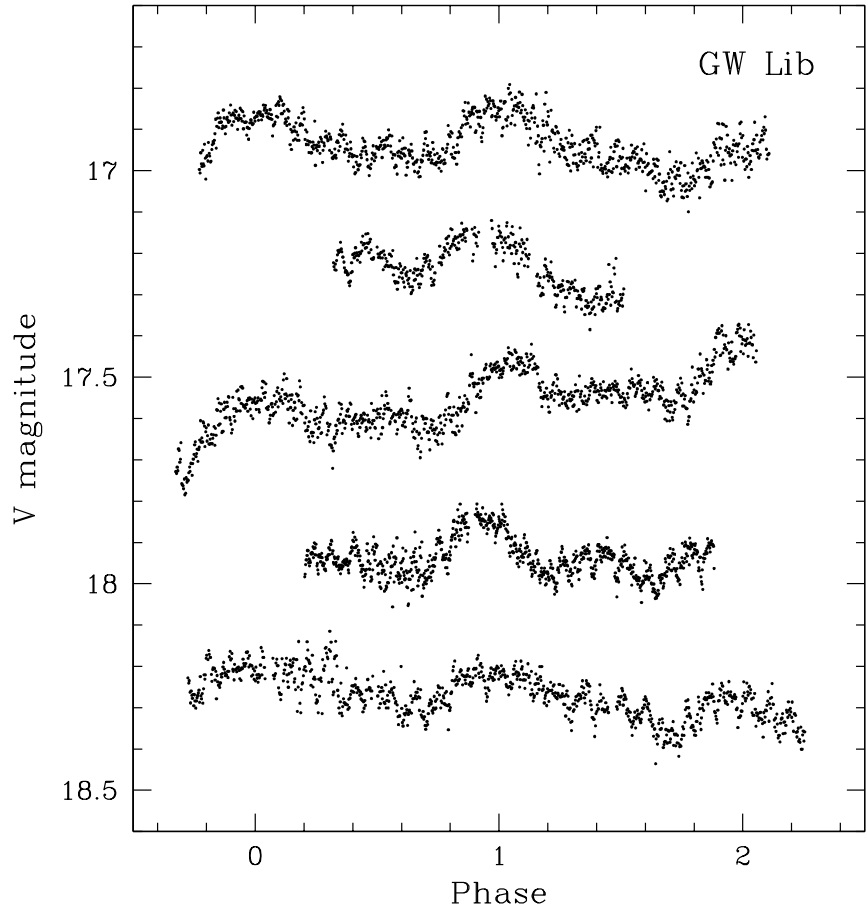}}
 \caption[]{Light curves of GW Librae, phased according to the ephemeris of Eqn.~1. The upper
curve is at the correct brightness. The others (in chronological order from top to bottom) have
been displaced vertically by arbitrary amounts for display purposes.}
 \label{fig1}
\end{figure}

\begin{figure}
\centering
 \resizebox{\hsize}{!}{\includegraphics{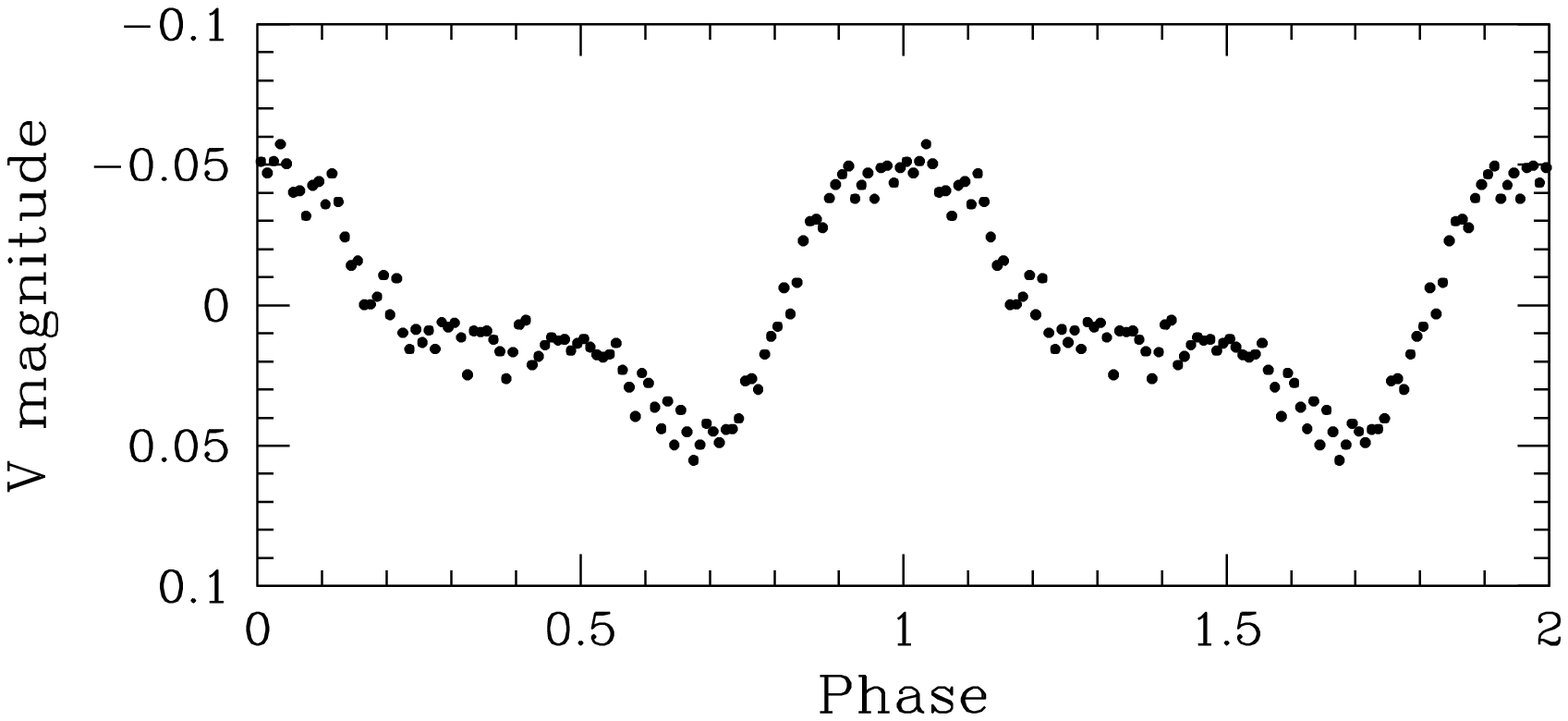}}
 \caption[]{Mean light curve of GW Librae.}
 \label{fig2}
\end{figure}

 The Fourier amplitude spectrum of the entire data set (with the means, and linear 
trends subtracted for each night) is shown in Figure 3. The dominant peaks are at 7526 s 
and 7780 s (with uncertainties, due to noise, of a few seconds), but there is also 
significant amplitude in the vicinity of their first harmonic (and some evidence for a weak 
second harmonic). Prewhitening at either of the dominant periods removes most of the 
power in the region - the variable amplitude of the modulation probably causes the 
remaining power. The highest peaks at the first harmonic are at 3759 s and 3822 s. For 
both these, and the double peak at the fundamental, the separation of the peaks is caused 
by a 1/2.6 d$^{-1}$ alias that is present in the window pattern of the Fourier transform. The fact 
that the mean profile of the modulation (Fig.~2) contains an obvious first harmonic 
enables us to choose the 7526, 3759 s pair (marked by diagonal bars in Fig.~3) as the correct 
fundamental and its harmonic -- the other pairings of frequencies do not have the 
necessary factor of two relationship.
    
 From the fundamental period in the Fourier transform, we find the time of maximum light
is given by the following ephemeris:

\begin{equation}
{\rm HJD_{max}} = 2452048.42488 + 0.\!\!^{\rm d}08711 \, {\rm E}
\end{equation}

 The modulation that we have found in GW Lib has therefore a doubled humped 
profile with a period of 7526 s = 125.4 min = 2.091 h, which persisted over a time of 
nearly two weeks. This neither equals nor bears any obvious simple relationship to the 
76.78 min spectroscopic period found by Thorstensen et al.~(2001). Prewhitening of our 
observations at the fundamental and first harmonic of the modulation does not reveal any 
significant signal near to the 76.78 min period.
   
 The mean brightness of GW Lib during our 2001 observations was not detectably 
different (less than $\sim$ 0.1 mag difference) from its brightness in 1997 and 1998. However, 
our light curves (Fig.~1) show evidence on some nights of slow systematic variations of 
0.1 -- 0.2 mag about the mean.

\begin{figure}
\centering
 \resizebox{\hsize}{!}{\includegraphics{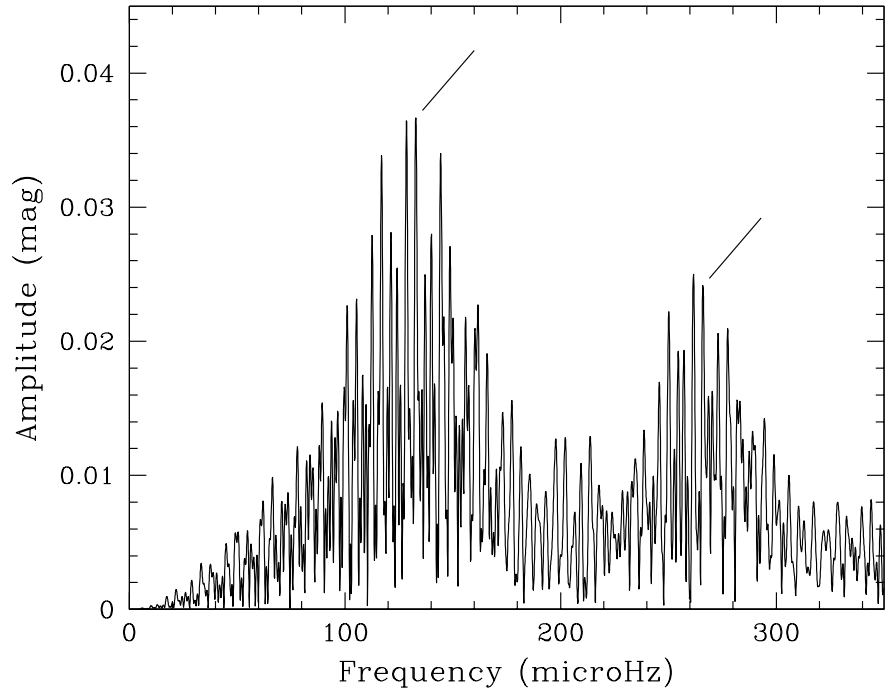}}
 \caption[]{Fourier spectrum of GW Librae. The fundamental and first harmonic of the 2.091 h period
are marked by diagonal bars.}
 \label{fig3}
\end{figure}

\section{Discussion}

The essential issues that GW Lib raises are (a) the appearance of a strong periodic 
brightness modulation where there was none before and (b) the period itself, which is 
different from the spectroscopic value.
  
 There are examples of orbital modulations, caused by the bright spot being obscured by 
the disc, which have disappeared for extensive lengths of time (e.g., HT Cas: Patterson 
1981; Horne, Wood \& Stiening 1991) and which may be the result of the disc becoming 
optically thin. The long outburst interval in GW Lib is characteristic of SU UMa stars 
with a period around 80 min (Warner 1995b), which have very low mass ratios and 
consequently are prone to develop elliptical accretion discs and their associated 
superhumps. The latter arise from stresses in the disc caused by the orbital passage of the 
secondary star, and are thus visible even in very low inclination systems. It could be, 
therefore, that the periodic modulation that we see in GW Lib is the result of the 
development of superhumps -- the double hump profiles are consistent with this. If this 
is the case, then the orbital period of GW Lib would be a few percent shorter than the 
2.09 h that we measure.
     
 On the other hand, there is no doubt that there is a clear periodic signal in the 
spectroscopy, with Szkody, Desai \& Hoard (2000) and Thorstensen et al. (2001) both 
detecting signals (but the latter using more extensive and better distributed observations). 
Both spectroscopic and photometric periods can, in various systems, be identified as the 
orbital period or the spin period of one of the components. There is no evidence in GW 
Lib (from the spectrum) of any strong magnetic field, so the two periods are unlikely to 
be orbital and white dwarf rotation, as they often are in intermediate polars (e.g. 98 min 
and 67 min respectively in EX Hya). The ratio of photometric to spectroscopic periods is 
only 1.64, and even if the beat between the two periods is thought to be relevant, the ratio 
of beat and spectroscopic periods is only 3.27. These low values appear to rule out any 
disc precession as the possible generator of the photometric period.
  
 There is evidence that a similar dilemma occurs in another system. FS Aur has a 
spectroscopic period of 85.7 min (Thorstensen et al. 1996), which has also been seen by 
Neustroev (2001), but the latter finds in addition photometric variations with range of 
about 0.25 mag and a possible period around 3 h. This period needs to be confirmed by 
more extensive photometry; there is no sign of the spectroscopic period in the existing 
photometry.
  
 Another anomalous system is V2051 Oph, in which the orbital period is definitively 
established from eclipses as 89.9 min. Warner \& O'Donoghue (1987), from photometry 
at quiescence, found a brightness modulation with range $\sim$ 0.15 mag and period $\sim$ 16\,500 
s (274 min) which was present for several nights during two observing runs a month 
apart. The ratio of periods in this case is 3.05.
   
 We do not have a physical explanation for the longer periods in GW Lib, FS Aur and 
V2501 Oph. Long term photometric studies, to examine the stability of the periods, and 
simultaneous photometry and spectroscopy may assist in developing a model.

\acknowledgements
    This research was supported by funds made available by the University of Cape Town. 
We are grateful to John Thorstensen and Joe Patterson for helpful comments.

\end{article}
\end{document}